\documentclass[twocolumn]{aastex62}

\usepackage{epsfig,psfig}
\usepackage{epstopdf}            
\usepackage{graphicx,color}        
\usepackage{color}            
\usepackage{psfig}
\usepackage{subfloat,subfigure}
\usepackage{auto-pst-pdf}
\usepackage{natbib}
\usepackage{hyperref} 

\hypersetup{linkcolor=red,citecolor=blue,filecolor=cyan,urlcolor=magenta}

\newcommand{\sdssfortythree}{SDSS J004305.27+194859.20~}
\newcommand{\sdsssixfortyeight}{SDSS J064813.33+323105.2~}

\newcommand{\sdssninetwentyone}{SDSS J092157.27+503404.7~}
\newcommand{\sdssseventeenthirty}{SDSS J173025.57+414334.7~}

\newcommand{\sdsstwentythree}{SDSS J231923.85+191715.4~}

\newcommand{\sdssnseventeenthirty}{SDSS J1730+4143~}

\graphicspath{{./}{figures/}}

\received{}
\revised{}
\accepted{}
\submitjournal{ApJ}

\shorttitle{New $r$-Process-Enhanced Stars}
\shortauthors{Bandyopadhyay et al.}

\begin{document}

\title{Abundance Analysis of New $r$-Process-Enhanced Stars from the HESP-GOMPA Survey}

\correspondingauthor{Avrajit Bandyopadhyay}
\email{avrajit.india@gmail.com, avrajit@iiap.res.in}

\author{Avrajit Bandyopadhyay}
\affiliation{Indian Institute of Astrophysics, Bangalore, India}

\author{Thirupathi Sivarani}
\affiliation{Indian Institute of Astrophysics, Bangalore, India}

\author{Timothy C. Beers}
\affiliation{Department of Physics and JINA Center for the Evolution
of the Elements, University of Notre Dame, Notre Dame, IN, 46656, USA}

\begin{abstract}

 We present a study on the detailed chemical abundances of five new relatively bright $r$-process-enhanced stars that were initially observed as part of the SDSS/MARVELS pre-survey.  These stars were selected, on the basis of  their metallicities and carbon abundances, among a total of 60 stars,
for high-resolution spectroscopic follow-up as part of the HESP-GOMPA
 survey (Hanle Echelle SPectrograph -- Galactic survey Of Metal Poor stArs). Here we discuss the three new $r$-I and two new $r$-II stars  found in this survey. We have carried out a detailed abundance analysis  for each of these stars, at a resolving power of $R \sim 30,000$, and compare our results to the existing literature.  We could measure three of the first  $r$-process-peak elements (Sr, Y and Zr) in all five stars, while Ba, Ce, Nd, Sm, Eu, and Dy could be detected among the second $r$-process-peak elements. Thorium could also be detected in one of the targets, which is found to be  an actinide-boost star.  We have carried out a comparative study among the sub-populations of the $r$-process-enhanced stars and other  stars of the Milky Way halo population to constrain the origin of this class of objects. These bright $r$-process-enhanced stars provide an excellent opportunity to study the nucleosynthesis history of this population in great detail, and shed light on their chemical-enrichment histories. 
 
\end{abstract}

\keywords{Galaxy: halo ---  stars : abundances ---  stars : Population II ---stars : individual --- nucleosynthesis }

\section{Introduction} \label{sec:intro}

Detection and measurement of the chemical abundances of the elements in 
very metal-poor stars provides a wealth of information regarding their star-formation history and the prevalent astrophysical conditions in the early epochs of the Galaxy \citep{beers2005, Frebelandnorris, frebelrev15}. The low iron content in these stars already indicates that their natal clouds  received contributions from very few previous generations of stars; in some cases this pollution could be from a single star (e.g., \citealt{placco2014}). Among the most important elements for tracing the nature of early-generation stars and the early evolution of the Galaxy are those formed by the rapid neutron-capture process (see, e.g., \citealt{sneden1996,christ2004,barklem2005,freb2007,roederer2014,placco17,hansen18rpa,holmbeck18,sakari18rpa}).  The recent review by \citet{cowan2020} provides an excellent summary of current understanding. 


The observed enhancements in the neutron-capture elements  for a subset of ancient, low-mass  ($0.57 M_{\odot}  \leq M \leq 0.8 M_{\odot}$)\citep{fubressan}, very metal-poor (VMP; [Fe/H] $\le -2.0$) stars in the halo  indicates that they have been produced through a variety of nucleosynthetic pathways, including the $s$-process, the $i$-process, and the $r$-process.  Enrichment in the $s$-process elements, and possibly the $i$-process \citep{hampel2016, den2017} elements, is attributed to mass transfer from a relatively more-massive binary companion during its AGB phase. \citet{lucatello2006}, and many others since, including, in particular \cite{hansen2016b}, have shown that the great majority (perhaps all) of the stars that exhibit an $s$-process abundance pattern among their heavy elements,  along with an enhancement in carbon, possess an unseen binary companion. Insufficient data presently exists for similarly strong claims to be made for the origin of stars with $i$-process abundance patterns, but a number of binaries are known among such stars.  On the other hand, the preponderance of evidence, including the absence of significant numbers of binaries relative to other ``normal" VMP stars 
(e.g., \citealt{hansen2016a}) suggests that stars with $r$-process abundance patterns were born from gas that had previously been polluted by a progenitor object (and were not the result of mass transfer from an evolved companion), the nature of which is still an active area of research.  

Production of the $r$-process nuclides requires explosive conditions, capable of producing a large fluence of neutrons with a specific range of entropy \citep{argast2004}. The primary site(s) suggested for the production of $r$-process elements are core-collapse supernovae, magneto-rotationally jet-driven supernovae, neutron star mergers, neutron star / black hole mergers, and collapsars (e.g., \citealt{lattimer74,rosswog2014,lippuner2017,siegel2019}). 
The pattern of $r$-process enhancement at the  second and third $r$-process peaks is (mostly) universal and robust among halo stars with these signatures. The stars with chemical imprints of enrichment by the $r$-process exhibit similar patterns as the scaled-Solar $r$-process residuals, at least for elements heavier than Ba. Among the lighter $r$-process elements (e.g., Sr, Y, Zr),   discrepancies have been noted by many observers \citep{barklem2005,honda2006,roederer2010,camilla2012,yong2013}. Current understanding is that the overall $r$-process-enhancement patterns in stars could be the result of multiple contributions from different kinds of progenitors, rather than single-progenitor production of the observed pattern.  A weak $r$-process (also known as the limited $r$-process) could be responsible for the production of the lighter elements, but lacked sufficient neutron flux to produce the heavier elements \citep{truran2002,frebelrev18}\footnote{A mechanism called the "Lighter Element Primary Process", or LEPP, was also proposed by \citet{travaglio2004}, which preferentially produces the lighter elements in the early Galaxy.}, whereas the main $r$-process could be responsible for the production of the second $r$-process-peak elements and beyond. 

Although a great deal of attention has been (appropriately) focused on the direct demonstration that at least one suggested site, neutron star mergers, can be associated with the production of heavy $r$-process elements \citep{freiburghaus1999, tsuji1, thielemann2017, cote2018, kenta2018}, there continues to be a need for intense studies of the range of observed behavior of the $r$-process patterns, from the lighter $r$-process elements to the actinides (e.g., Th and U),  in dwarf galaxies (in particular the ultra-faint dwarfs such as  Reticulum II and Tucana III -- see, e.g., \citealt{ji2016apj}, \citealt{roederer2016ret}, \citealt{frebel2019ret}, \citealt{marshall2019}, and \citealt{reichert2020}), as well as in halo field stars.  The VMP $r$-process-enhanced (hereafter, RPE) halo stars, in particular the brightest of these, are the ideal targets to constrain the likely astrophysical sites and environments in which $r$-process-element enhancements were produced during the earliest epochs of star formation in the Galaxy.

\section{Observations and Analysis} 

High-resolution ($R \sim 30,000$) spectroscopic observations of our
five program stars,  covering the wavelength range 3600-10800\,{\AA}, were carried out as part of the GOMPA (Galactic
survey Of bright Metal Poor stArs) survey, using the Hanle Echelle
Spectrograph on the 2-m Himalayan Chandra telescope (HCT) at the
Indian Astronomical Observatory (IAO). The full set of 60 bright VMP targets were selected from the medium-resolution ($R \sim 2000$) spectroscopic pre-survey for MARVELS, which was carried out as part of SDSS-III. This offered the
chance to identify bright halo stars which could be studied at high
spectral resolution using moderate-aperture telescopes.  As has been recognized for some time (e.g., \citealt{barklem2005}), moderate to highly RPE stars only constitute a small fraction of the halo VMP stars, hence the discovery of  even a few new examples is a valuable step forward. Observational details for the newly detected $r$-process-rich stars from our study are listed in Table 1. 

Radial velocities for our stars were computed by cross-correlating the observed spectra with a synthetic template of similar low metallicity and effective temperature. Photometeric, as well as spectroscopic, data have been used to estimate the stellar atmospheric parameters for these stars. The $T_{\rm eff}$, $\log (g)$, [Fe/H],
microturbulent velocity, $\xi$, and the abundances of individual elements
present in each spectrum were determined using standard procedures, as summarized below. The linelist  for these analyses is compiled from \citet{susmitha}, \citet{bandyopadhyay}, and the Kurucz database \footnote{http://kurucz.harvard.edu/linelists.html}.

Stellar parameters were determined using the same methods as described in \citet{bandyopadhyay} and \citet{bandyopadhyay2}. Briefly, photometric temperatures
were obtained using the available data in the literature and the
standard $T_{\rm eff}$-color relations derived by \citet{alonso1996} and
\citet{alonso1999}. VOSA, the
online spectral energy distribution (SED) fitter \citep{bayo2008vosa}, was also employed to derive the temperatures. Spectroscopically, $T_{\rm eff}$ estimates have been derived by demanding that there be no trend of Fe~I line
abundances with excitation potential, as well as by fitting of the
$H_{\alpha}$ profiles.

 An iterative method was employed to adopt the final values of the stellar atmospheric parameters.  The first estimate of temperature was based on the $V - K$ color, which is not expected to be strongly affected by metallicity. We then fixed the final value by fitting the wings of the H-$\alpha$ region, as they are highly sensitive to small changes in temperature, particularly for the temperature range of the program stars. The strengths of the Fe~I lines were employed to check for consistency with the adopted temperature. All the estimates for $T_{\rm eff}$ are tabulated in Table 1, which also includes values from $Gaia$ and the SED. \footnote{We note a significant departure for \sdssninetwentyone and \sdssseventeenthirty for the temperature obtained from their SEDs, due to the poor quality of the fit between the models and observed data in VOSA.}

 First estimates of surface gravity, $\log (g)$, were determined by the usual technique that demands equality of the iron abundances derived for the neutral (Fe~I) lines and singly
ionized (Fe~II) lines by fixing the temperature obtained previously. Fine tuning was accomplished by fitting the wings of the Mg~I lines, which are sensitive to small variations of $\log (g)$. Parallaxes from $Gaia$  have also been employed to
derive the $\log (g)$ for individual stars, to check for consistency with the adopted parameters.

We also ensured that the metallicity of the model atmosphere was the same as obtained from analysis of the Fe~I lines. The microturbulent velocity was obtained by varying it to achieve zero trend between the Fe~I line abundances and their reduced equivalent widths. Finally, all the adopted atmospheric parameters for each star were checked for convergence. The final stellar parameters for our program stars are listed in Table 2.

Abundance estimates for the various elements present in the program stars were derived by employing one-dimensional local thermodynamic equilibrium (LTE) stellar atmospheric models (ATLAS9; \citealt{castellikurucz}) and version 12 of the spectral synthesis code TURBOSPECTRUM \citep{alvarezplez1998}. Both spectrum synthesis and equivalent-width measurements were employed to derive the abundances.

 Errors in the derived abundances primarily depend on the signal-to-noise ratio (SNR) of the observed spectrum and deviations in the values of the adopted stellar parameters. We have used the relation given by \cite{cayrel1988} to calculate the uncertainty in the abundances due to the SNR. Uncertainties due to possible temperature and $\log (g)$ deviations were derived using two different model spectra, the first differing in temperature by $\sim$150 K and the second one deviating in $\log (g)$ by 0.25 dex. The final values of the abundance errors were obtained by adding the uncertainties arising from all three sources in quadrature. However, the errors in the relative abundance ratios are less sensitive to the errors in the model parameters, and  mainly depend on the SNR.

 The abundances of a given element can be referred to an absolute scale, relative
to the assumed number density of hydrogen atoms, defined by  log $\epsilon$(A) = $\log(N_{\rm A}/N_{\rm H}) + 12.0$, 
where A is taken to represent the element under consideration. In the standard nomenclature, the elemental-abundance ratios for two elements in a star are defined relative to the respective abundances in the Sun as \mbox{[A/B]} $ = \log(N_{\rm A}/N_{\rm B}) - \log(N_{\rm A}/N_{\rm B})_\odot$, where $N_{\rm {A}}$ and $N_{\rm {B}}$ are the number densities of atoms of element A and B, respectively. Due to the large number of Fe lines in a star, iron is often chosen as the standard reference element; results are listed in Tables 4 - 8, as described below.

\begin{table*}
\centering
\begin{center}
\caption{Estimates of Effective Temperature (K) for the Program Stars}
\begin{tabular}{ccccccccccrrrrrrr}
\hline\hline
Object &Short Name &H-$\alpha$ &Fe~I &$V-K$ &$J-H$ &$J-K$ &$Gaia$ &SED\\
\hline

SDSS J004305.27+194859.2 &SDSS J0043+1948 &4500 &4600 &4608 &4729 &4679 &4877 &4750 \\
SDSS J064813.33+323105.2 &SDSS J0648+3231 &4800 &4700 &4620 &4789 &4779 &5016 &4750 \\
SDSS J092157.27+503404.7 &SDSS J0921+5034 &4800 &5200 &4984 &4704 &5024 &5450 &6000 \\
SDSS J173025.57+414334.7 &SDSS J1730+4143 &4900 &4900 &4922 &4740 &4948 &5050 &5750 \\
SDSS J231923.85+191715.4 &SDSS J2319+1917 &4500 &4600 &4378 &4364 &4384 &4644 &4500 \\

\hline
\end{tabular}
\end{center}
\end{table*}

\begin{table*}
\centering
\begin{center}
\caption{Observation Details and Adopted Stellar Atmospheric Parameters for the Program Stars}
\begin{tabular}{ccccccccccrrrrrrr}
\hline\hline
Object &Short Name &Obs.Time &SNR &$V$ &Rad. Vel. &$T_{\rm eff}$ &$\log (g)$ & $\xi$ &[Fe$/$H]\\
       &           & (secs)  &    &    & (km~s$^{-1}$) & (K)      & (cgs)  & (km~s$^{-1}$) & \\ \\
\hline
SDSS J004305.27+194859.2 &SDSS J0043+1948 &10800 &130.2 &9.90 &$-$196.5 &4500 &1.50 &1.80 &$-$2.25 \\
SDSS J064813.33+323105.2 &SDSS J0648+3231 &7200 &117.8 &9.92 &135.5 &4800 &1.70 &1.80 &$-$2.35\\
SDSS J092157.27+503404.7 &SDSS J0921+5034 &7200 &50.9 &11.75 &$-$130.5 &4800 &1.75 &1.50 &$-$2.65\\
SDSS J173025.57+414334.7 &SDSS J1730+4143 &7200 &58.7 &12.03 &$-$133.0 &4900 &2.50 &1.75 &$-$2.85\\
SDSS J231923.85+191715.4 &SDSS J2319+1917 &8100 &77.0 &11.60 &$-$250.5 &4500 &1.25 &1.50 &$-$2.10\\

\hline
\end{tabular}
\end{center}
\end{table*}

\section{Abundances}

\subsection{Light and Iron-peak Elements}

Abundances for carbon have been derived from the molecular CH $G$-band around the 4313\,{\AA} region, using the method of spectrum synthesis. The spectral fitting for the $G$-band region is shown for three of the stars in Figure 1.  The dots denote the observed spectra, while the colored lines indicate the synthetic spectra.  The best fit for each target star is marked in red, while the blue and black lines indicate deviations of $\pm$\,0.50 dex. Nitrogen could be measured for only one of the stars, \sdssnseventeenthirty. The CN molecular band at 3883\,{\AA} was used to derive the abundances of N by iteratively changing the N abundances while keeping the C abundances constant at the values derived from the $G$-band synthesis. Measurements of the CN-band region are difficult using HESP spectra, owing to the poor signal-to-noise in that region.

\begin{figure*}[!htbp]
\centering
\includegraphics[width=2.0\columnwidth]{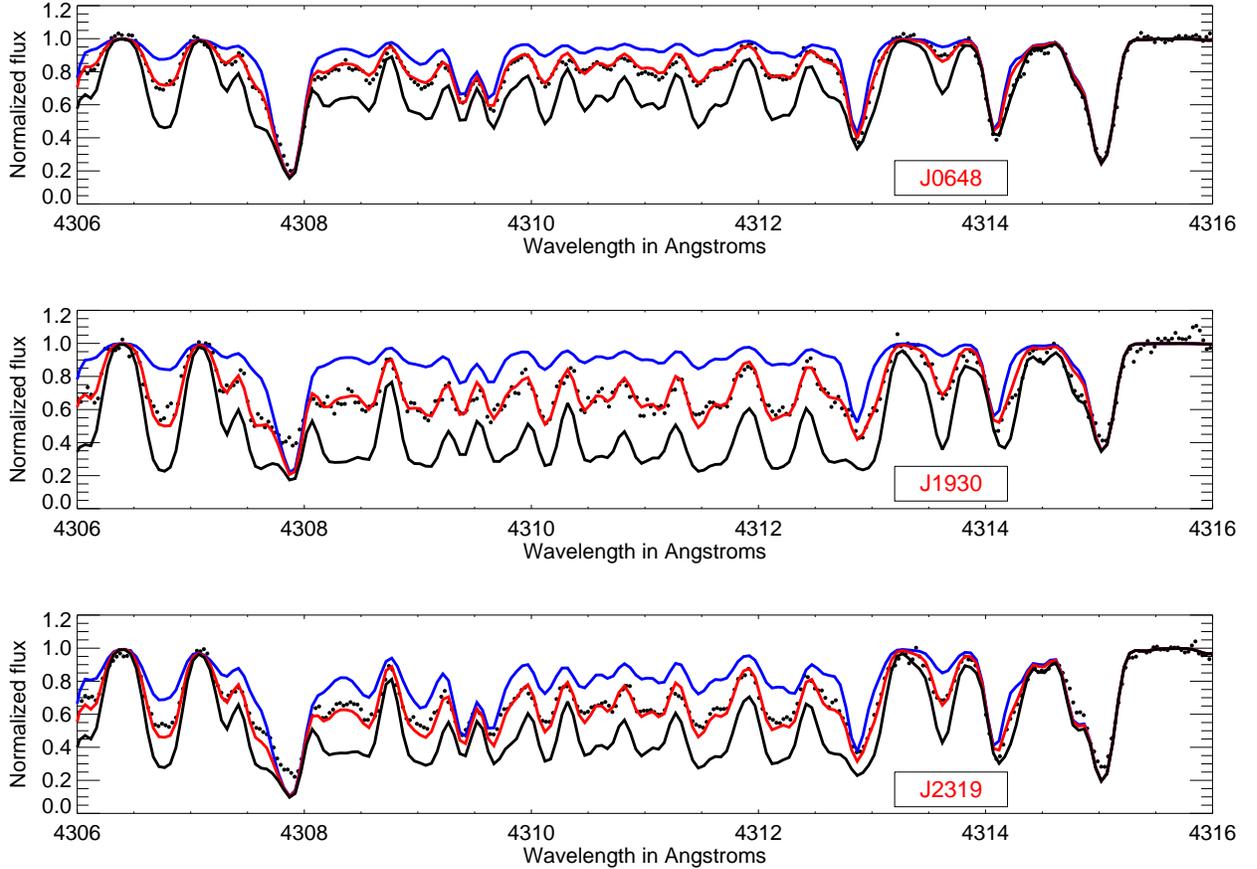}
\caption{Fits for the molecular carbon CH $G$-band for three of our RPE stars. The dots indicate the observed spectra, and the red solid lines mark the best spectral fit. The blue  and black solid lines indicate $\pm$\,0.50 dex around the best fits.} \label{c5f1}
\end{figure*}

Among the light elements, abundances for Na, Mg, Al, and Ca could be derived for all of the stars. A mixture of equivalent-width analysis and spectral synthesis was used for measuring the abundances. Spectrum synthesis was primarily used for the weaker lines, whereas the equivalent-width method was employed for the relatively strong and clean lines. NLTE corrections were applied, on a case-by-case basis, for Na and Al, as discussed by \citet{baumuller} and \citet{andrievskyna,andrievskyal}.  Enhancement in the $\alpha$-elements of about +0.4 dex, as expected for typical halo stars, was noted for all the objects except the most metal rich star of our sample SDSS~J2319+1917, which indicates that it might have undergone a different star-formation history. The space velocities ($u$, $v$, and $w$) have also been determined for SDSS~J2319+1917, and are found to be consistent with space velocities for halo stars \citep{kinman2007}. Thus it could be among the rare $\alpha$-element poor RPE stars in the halo at this metallicity. The space velocities of all the stars in our sample are given in table 9. A detailed study on space velocities of RPE stars can be obtained from \cite{roederer18_spvel}.

Abundances could be derived for the Fe-peak elements, such as Sc, Cr, Mn, Co and Ni, by the method of equivalent-width analysis. Additionally, Cu and Zn could also be measured for a few of the program stars. The derived abundances are similar to other normal stars of similar metallicity; they tend to track the Fe content of the star.  

Results for all five program stars are provided in Tables 4 -- 8.

\subsection{Neutron-Capture Elements}

For the five stars studied here, we could obtain abundances for up to 11 neutron-capture elements. The primary factors that prohibited us from measuring a larger number of neutron-capture elements in these stars are twofold:
\begin{itemize}

\item Most of our VMP target stars have relatively higher metallicity, as seen in Table 2, and are generally quite cool, hence they exhibit stronger metal lines compared to RPE extremely metal-poor (EMP; [Fe/H] $\le -3.0$) stars with higher effective temperatures. These metallic lines cause blending, which creates difficulties in measuring the weak neutron-capture elements.

\item The poor SNR at the blue end of the spectra. The bulk of the spectral lines of the key neutron-capture elements fall blueward of the 4000\,{\AA} region, where the quality of the signal in our spectra becomes degraded, and no meaningful abundances could be derived.

\end{itemize}
The abundances for the elements Sr, Y, Zr, Ba, Ce, Nd, Sm, Eu, Dy, and Th were measured using spectral synthesis, accounting for hyperfine transitions \citep{mcwilliam1998,cui_heres}. 
Strontium is usually measured from two lines -- 4077\,{\AA} and 4215\,{\AA}. The 4077\,{\AA} line is in some instances too strong for reliable measurement; in those cases only one line at 4215\,{\AA} is used to derive the abundance. Abundances for Ba are derived using four lines -- 4554\,{\AA}, 4934\,{\AA}, 5853\,{\AA}, and 6141\,{\AA}. NLTE corrections have also been taken into account for the strong  4077\,{\AA}, 4215\,{\AA}, and 4554\,{\AA} lines, following \citet{short2006}, who found only a small offset of +0.14 dex in the case of Ba, and even smaller offsets for Sr. Europium abundances could also be derived for all of the program stars; the line at 4129\,{\AA} was used to derive the Eu abundances. The Th abundance has been derived using the spectral line at 
4019\,{\AA}, which required special attention, as there are  very nearby lines of Nd and Co present. Thus, abundances of Nd and Co were determined using other lines in the spectrum, and were held constant, while the Th abundances were varied iteratively to obtain the best fit in the region.  In the end, we could only derive a Th abundance for one star, SDSS~J0043+1948.  The spectral synthesis for three of the most important elements to classify $r$-process-rich stars -- Eu, Sr, and Ba, are shown in Figure 2. The spectral fits for some of the other important lines for different target stars are shown in Figure 3.

\begin{figure*}[!htbp]
\centering
\includegraphics[width=2.0\columnwidth]{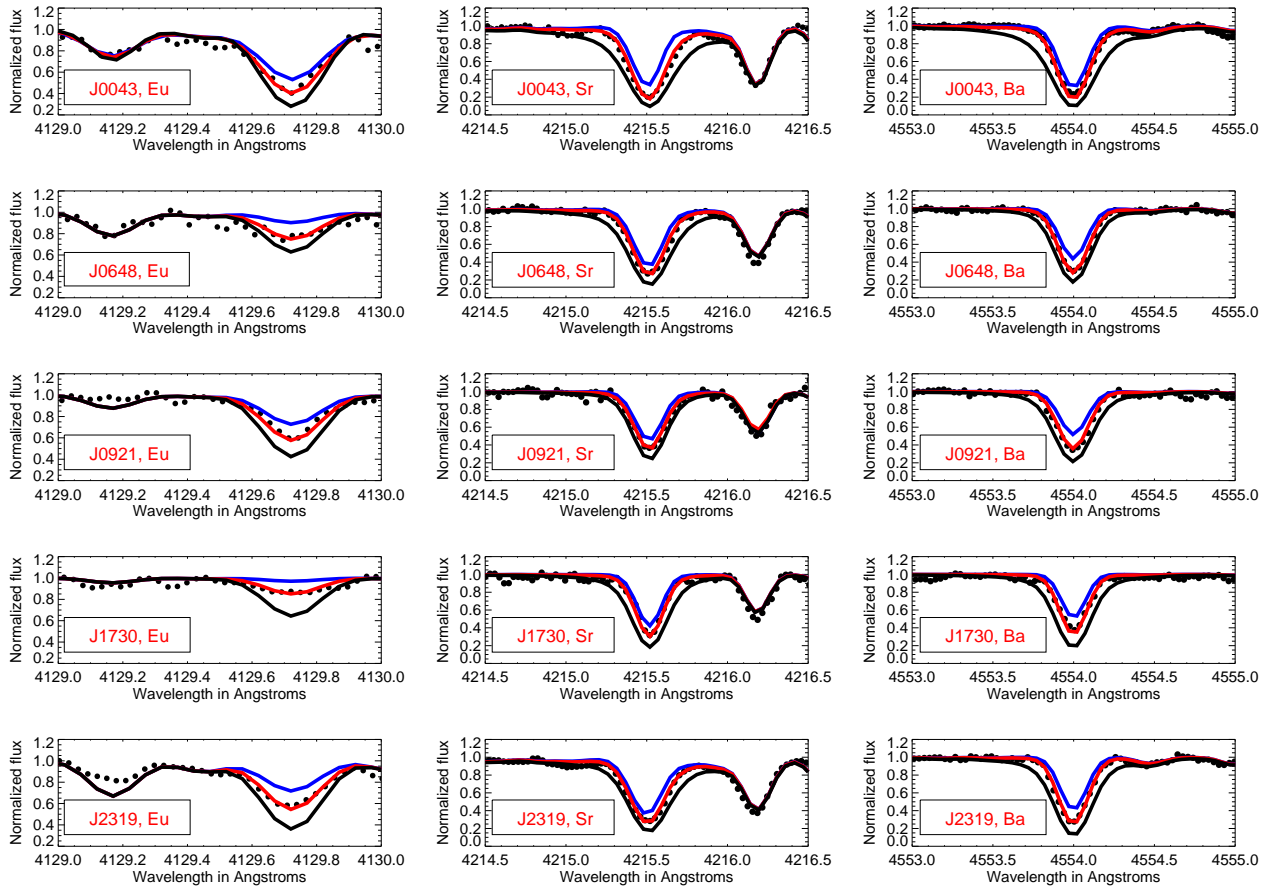}
\caption{Spectral fitting for the key $r$-process elements Eu, Sr, and Ba. Filled black circles represent the observed spectra, and red is used to indicate the 
best-fit synthetic spectra. The blue and black solid lines indicate abundances differing by $\pm$\,0.50 dex from the best-fit value.} \label{c5f2}
\end{figure*}

The element Dy could  only be measured for one of our program stars (SDSS~J2319+1917), while an upper limit could be obtained for two others (SDSS~J0043+1948 and SDSS~J0921+5034). Example syntheses for some of the other key $r$-process elements, such as Y, Zr, La, Ce, Nd, and Th are shown for different stars in Figure 3.

\begin{figure*}[!htbp]
\centering
\includegraphics[width=2.0\columnwidth]{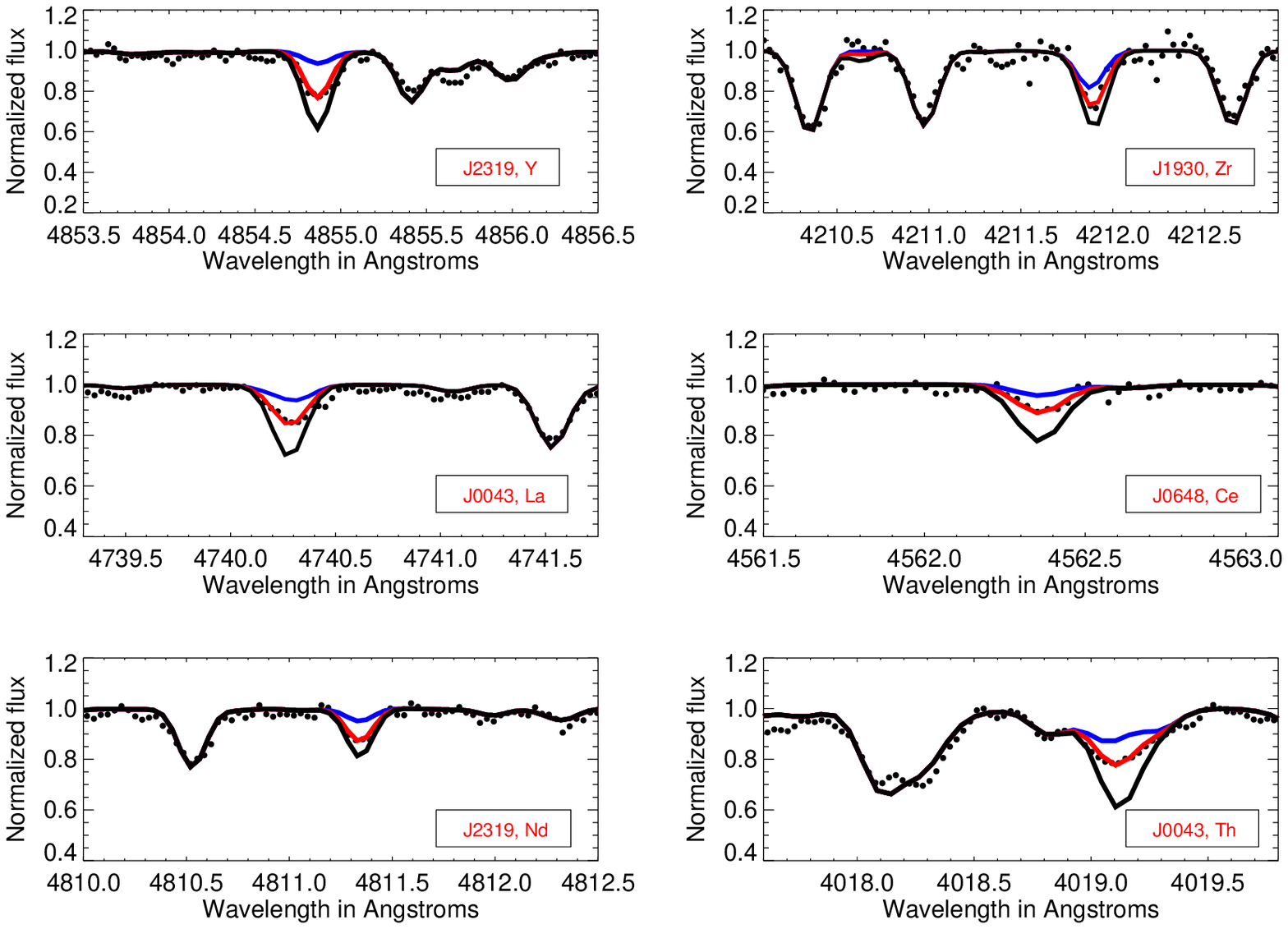}
\caption{Spectral fits for some of the important $r$-process elements for different stars. Black dots indicates the observed spectra and red lines mark the best-fit synthetic spectra. The blue and black solid lines indicate abundances differing by $\pm$\,0.50 dex from the best-fit value.} \label{figc5f3}
\end{figure*}

\section{Discussion}
\subsection{New Identifications of RPE Stars}

In this study, we have identified five new VMP halo stars with moderate-to-high enhancements in $r$-process-element abundances. Following the standard definitions in \citet{beers2005}, two of them are $r$-II stars ([Eu/Fe] $>$ +1.0 and [Ba/Eu] $<$ 0.0), while the other three are $r$-I stars (+0.3 $\leq$ [Eu/Fe] $\leq$ +1.0 and [Ba/Eu] $<$ 0.0), as shown in Table 3. The distribution over metallicity for the objects are from [Fe/H] = $-$2.85 to [Fe/H] = $-$2.10, spanning a range of 0.75 dex.  The metal-poor end of the metallicity distribution is more suited to finding such stars during the HESP-GOMPA survey, as the $r$-process signatures are not suppressed significantly by subsequent chemical evolution. 
At higher metallicities, only stars with the strongest $r$-process signatures ($r$-II stars) could be detected at our resolution, owing to the strong metal lines  found in the the bluer region of the spectrum, (e.g., J18024226-4404426, at [Fe/H] = $-$1.55 and [Eu/Fe] = +1.05; \citealt{hansen18rpa}, or J1124+4535 at [Fe/H] = $-$1.27, [Eu/Fe]= +1.10; \citealt{xing2019}).
The moderate spread in [Eu/Fe] in our sample, 0.80 dex, could be attributed to the contributions from several $r$-process events \citep{sneden2000, travaglio2004, hansen18rpa}, or, perhaps more likely, due to different levels of dilution associated with the baryonic masses of the natal clouds in which these stars formed \citep{jifrebelnature16, frebelrev18}. The spread of our sample is shown in Figure 4, with the five program stars marked with red diamonds. The data for the reference stars has been compiled from the first two data releases of the $R$-Process Alliance (RPA; \citealt{hansen18rpa}, and \citealt{sakari18rpa}), and a few other well-known $r$-process-rich stars. The dashed lines indicate the different levels of enhancements used to differentiate $r$-II, $r$-I, and limited-$r$ stars.

\begin{table*}
\begin{center}
\caption{Classifications of the Program Stars}
\begin{tabular}{ccccccccccr}
\hline\hline
Object &[Fe/H] &[C/Fe]$^o$ &$\Delta$ [C/Fe]$^e$ &[C/Fe]$_{corr}$ & [Sr/Fe] &[Ba/Fe] &[Eu/Fe] &[Ba/Eu] &[Sr/Ba] &Class \\
\hline

SDSS~J0043+1948    &$-$2.25 &$-$0.93 &$+$0.49    &$-$0.44    &$+$0.08    &$+$0.47    &$+$0.98    &$-$0.51    &$-$0.39     &$r$-I\\
SDSS~J0648+2321    &$-$2.35 &$-$0.58 &$+$0.29    &$-$0.29    &$+$0.27    &$+$0.35    &$+$0.68    &$-$0.33    &$-$0.08    &$r$-I\\
SDSS~J0921+5034   &$-$2.65 &$-$0.28 &$+$0.18   &$-$0.10    &$+$0.37    &$+$0.40    &$+$1.23    &$-$0.83    &$-$0.03    &$r$-II\\
SDSS~J1730+4143   &$-$2.85 &$-$0.08 &$+$0.01   &$-$0.07   &$+$0.52    &$+$0.40    &$+$1.08    &$-$0.68    &$+$0.12    &$r$-II\\
SDSS~J2319+1917   &$-$2.10 &$-$0.58 &$+$0.66   &$+$0.08    &$+$0.07    &$-$0.20    &$+$0.43    &$-$0.63     &$+$0.27    &$r$-I\\

\hline
\end{tabular}
\end{center}

    $^o$ Indicates the originally measured C abundances.
    \newline
    $^e$ Corrections due to evolutionary effects on the C abundances given by  \citet{placcocemp2014}.

\end{table*}

\begin{figure*}[!htbp]
\centering
\includegraphics[width=1.25\columnwidth]{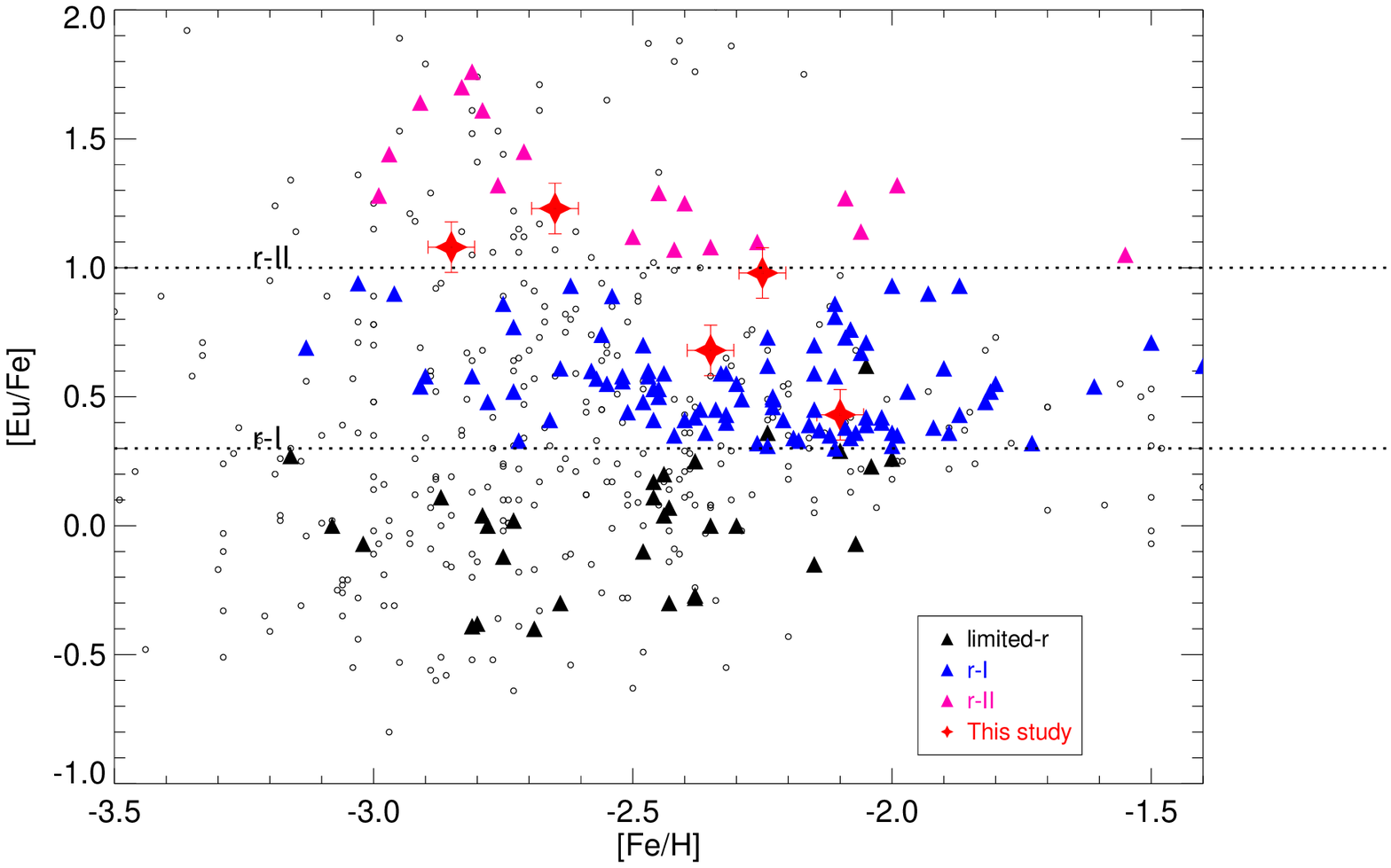}
\caption{Distribution of [Eu/Fe] for our program stars, as a function of [Fe/H], marked with filled red 
diamonds. Previously recognized $r$-I (blue triangles), $r$-II (pink triangles), and limited-$r$ 
(black triangles) are shown for comparison, compiled from the RPA papers (\citealt{placco17}, \citealt{cain18}, \citealt{hansen18rpa}, \citealt{holmbeck18}, \citealt{sakari18rpa}). 
The open circles indicate other stars with measured Eu abundances, as compiled from JINAbase \citep{jinabase} 
and SAGA databases \citep{sudasaga}, including \citet{mcwilliam1995,ryan1996,norris1997,burris2000,
johnson2002,cayrel2004,wako2005,preston2006aj,lai2008,roederer2010,iroederer2010,hollek2011,honda2011,
jcohen213,miho2013,placco2014,roederer2014,spite2014,tthansen2015,jacobson2015,li2015lamost}. 
The dashed lines correspond to the [Eu/Fe] levels used to distinguish the $r$-I and $r$-II stars.} 
\end{figure*}

\subsection{The Sub-Populations}

As discussed previously, enhancement in the neutron-capture elements in a metal-poor star could be caused by a number of nucleosynthetic pathways, including the $s$-, $i$-, and $r$-processes.  At low metallicity, Eu is a representative element of the $r$-process \citep{simmerer2004}, whereas Ba is predominantly produced by the $s$-process, but with a contribution from the $r$-process as well. The relative contribution of the $i$-process to most elements is still not well-understood.  Several  very and extremely metal-poor AGB stars produce large amounts of Pb through operation of the $s$-process \citep{sivarani2004}.  However, the predominant $s$-process contributions to the Galaxy begin at later times; $s$-process-enhanced tars are typically associated with AGB mass transfer, carbon enhancement, and the presence of a binary white dwarf companion. Lead is also one of the end products of the radioactive decay of U and Th after completion of the $r$-process, and could be a useful indicator as well, but we were unable to detect Pb in any of our spectra. Thus, [Eu/Ba] is often used to probe the relative contribution of one process to the others \citep{barklem2005, frebelrev15}.

\begin{figure*}[!htbp]
\centering
\includegraphics[width=1.25\columnwidth]{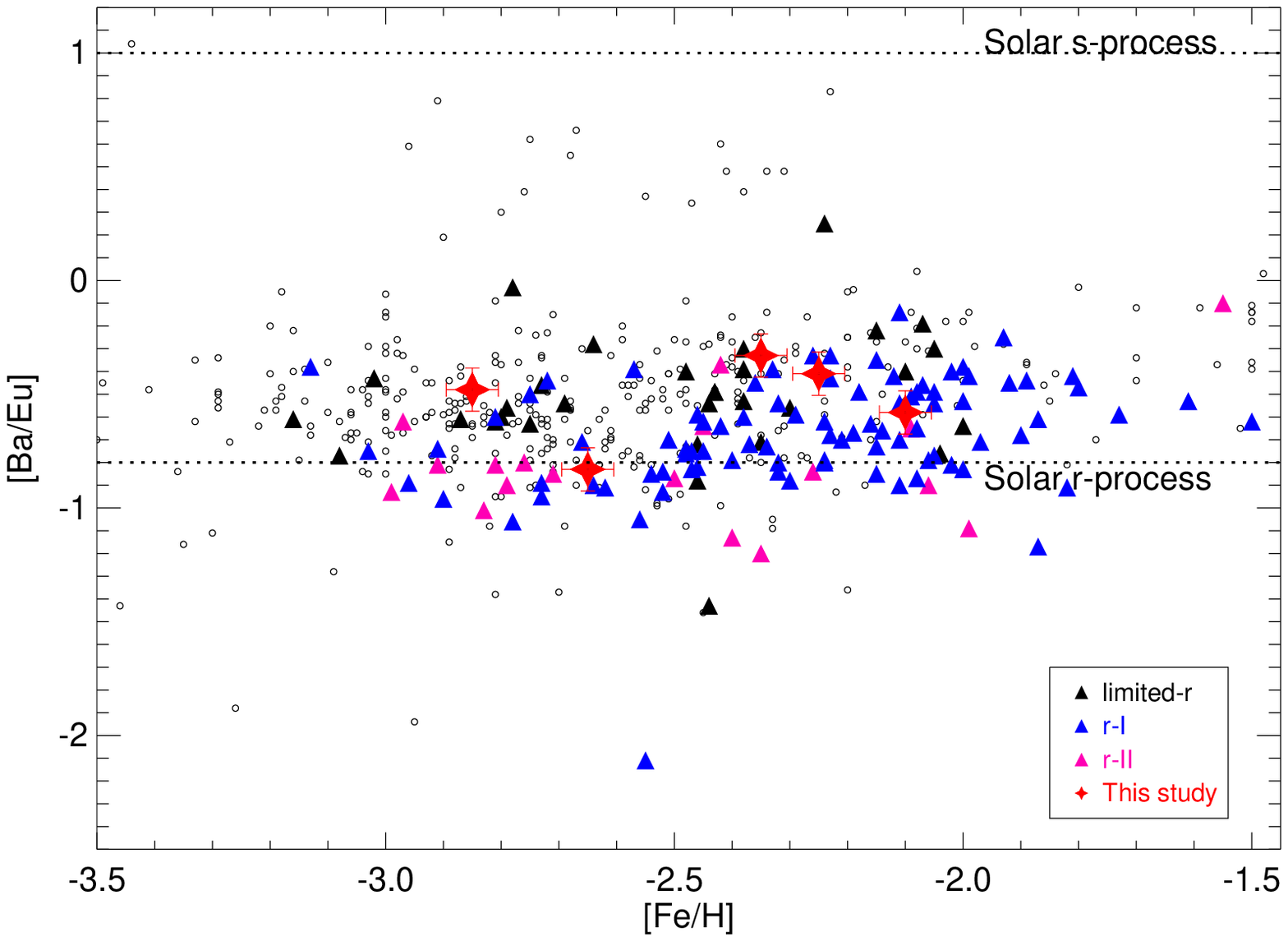}
\caption{Distribution of [Ba/Eu] for our program stars, as a function of [Fe/H], marked with filled red 
diamonds. Previously recognized $r$-I (blue triangles), $r$-II (pink triangles), and limited-$r$ 
(black triangles), are shown for comparison, compiled from RPA papers (\citealt{placco17}, \citealt{cain18}, 
\citealt{hansen18rpa}, \citealt{holmbeck18}, \citealt{sakari18rpa}). The open circles indicate other stars 
with measured [Ba/Eu] ratios, as compiled from the JINAbase \citep{jinabase} and SAGA databases 
\citep{sudasaga}, including \citet{mcwilliam1995,ryan1996,norris1997,burris2000,johnson2002,cayrel2004,
wako2005,preston2006aj,lai2008,roederer2010,iroederer2010,hollek2011,honda2011,jcohen213,miho2013,placco2014,
roederer2014,spite2014,tthansen2015,jacobson2015,li2015lamost}.
The dashed lines represent the levels of the Solar System $r$-process fraction ([Ba/Eu] = $-$0.80) and 
Solar System $s$-process fraction ([Ba/Eu] = +1.0) \citep{simmerer2004}.
} 
\end{figure*}

Following \citet{simmerer2004}, [Ba/Eu] $<$ $-$0.80 implies a pure $r$-process origin, whereas [Ba/Eu] $>$ +1.0 marks the Solar $s$-process. Figure 5 provides a plot of this ratio for the program and comparison stars previously shown in Figure 4.  As expected, all the known RPE stars exhibit [Ba/Eu] $<$ 0.0, indicating a higher contribution from the $r$-process in comparison to the $s$-process. However, only a few stars lie in the regime of pure $r$-process origin, with [Ba/Eu] $<$ $-$0.80. Among our sample, only one of the program stars represent a pure $r$-process origin.

A number of metal-poor halo stars are known to exhibit low abundances of the heavy $r$-process elements ([Eu/Fe] < +0.3), but 
higher abundances of the light $r$-process elements, such
Sr, relative to the heavy $r$-process
elements ([Sr/Ba] $>$ +0.5). These stars display the signature of an $r$-process that is limited by a smaller
flux of neutrons, and thus produces primarily lighter $r$-process elements  The origin of these elements is often attributed to a separate ``weak" $r$-process event (or the  Light Element Primary Process; LEPP, now referred to as the limited $r$-process) that preferentially produces the lighter $r$-process elements relative to the heavier ones \citep{frebelrev18}. HD~122563 is the canonical example of this group \citep{honda2006}.

 Figure 6 shows the ratio of [Sr/Ba] vs. [Eu/Fe] for the stars in our program and comparison samples with these abundances available. The limited-$r$ stars, marked in black, occupy the top-left corner of the plot. None of our program stars are found to belong to this category.  The ratio of [Sr/Ba] declines steadily as [Eu/Fe] increases, which again suggests different sites and mechanisms for the production of the lighter elements. Many studies have suggested that high-entropy neutrino winds from core-collapse supernovae could be responsible for the limited-$r$ stars \citep{woosleyhoffman92, kratz2007, arcones2011, wanajo2013, martinez6}.  As seen in the left-hand panel of Figure 6, the majority of the $r$-I and $r$-II stars are consistent with the main $r$-process ratio of [Sr/Ba]. The limited-$r$ stars exhibit a flatter trend with respect to  Eu, indicating a different origin for its production. To remove the dependence on metallicity, we have plotted [Sr/Ba] as a function of [Eu/H] in the right-hand panel, where a similar distribution is obtained. 

\begin{figure*}[!htbp]
\centering
\includegraphics[width=1.05\columnwidth]{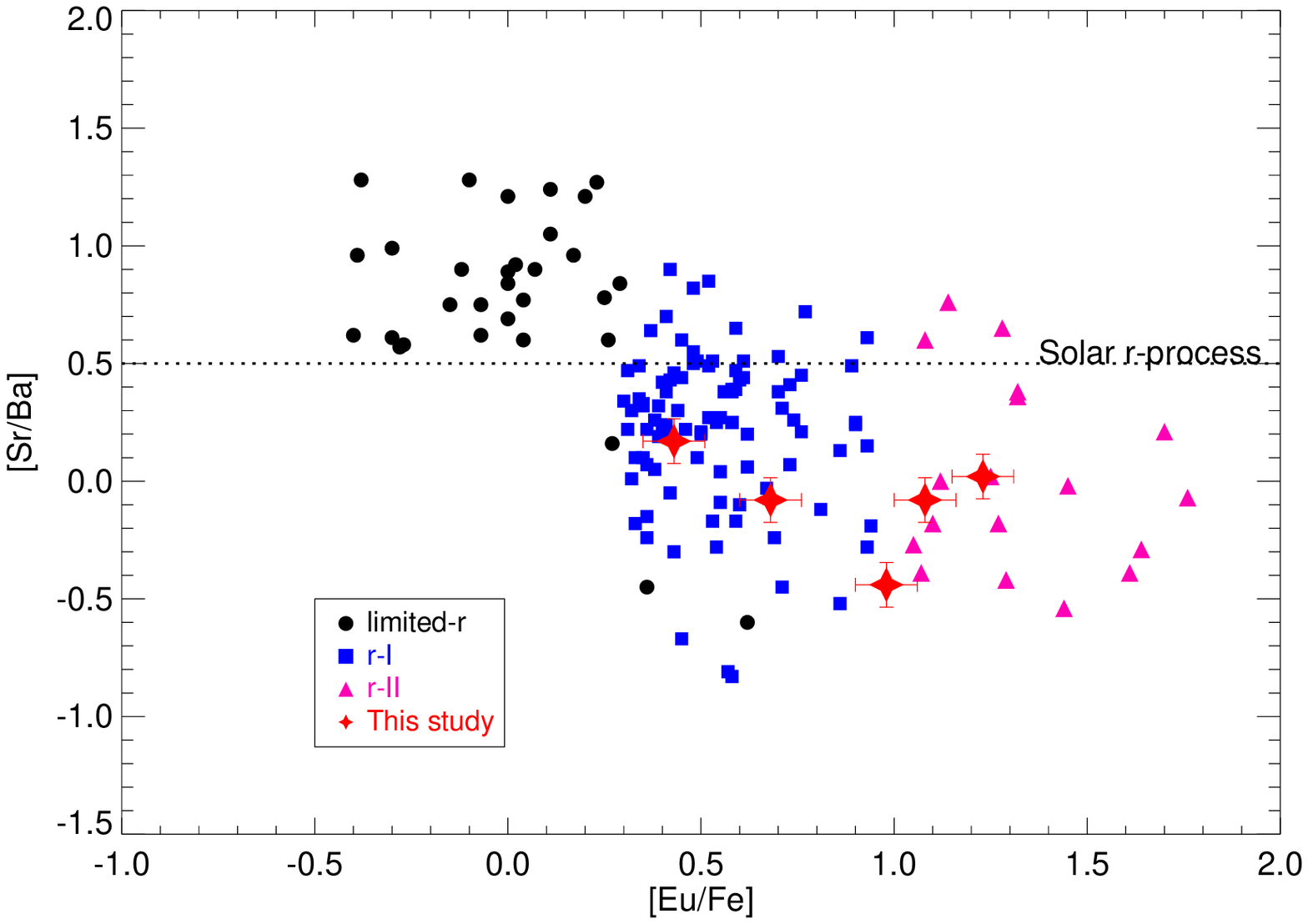}
\includegraphics[width=1.05\columnwidth]{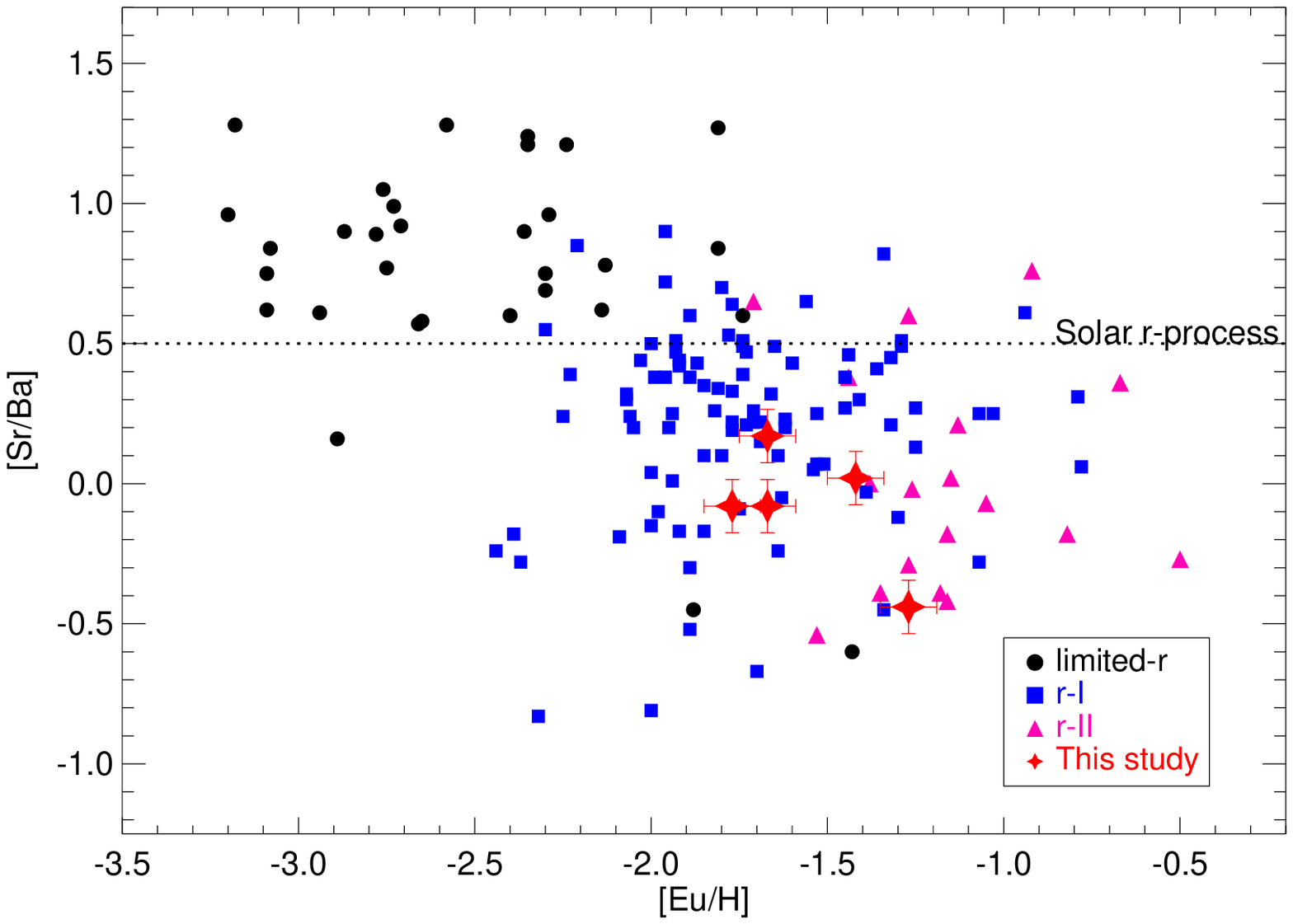}
\caption{Distribution  of [Sr/Ba] for our program stars, as a function of [Eu/Fe] and [Eu/H], marked with 
filled red diamonds. Previously recognized $r$-I (blue squares), $r$-II (pink triangles), and limited-$r$ 
(black circles) are shown for comparison, compiled from the RPA data releases \citep{hansen18rpa,sakari18rpa}.}
\end{figure*}

\begin{figure*}[!htbp]
\centering
\includegraphics[width=1.05\columnwidth]{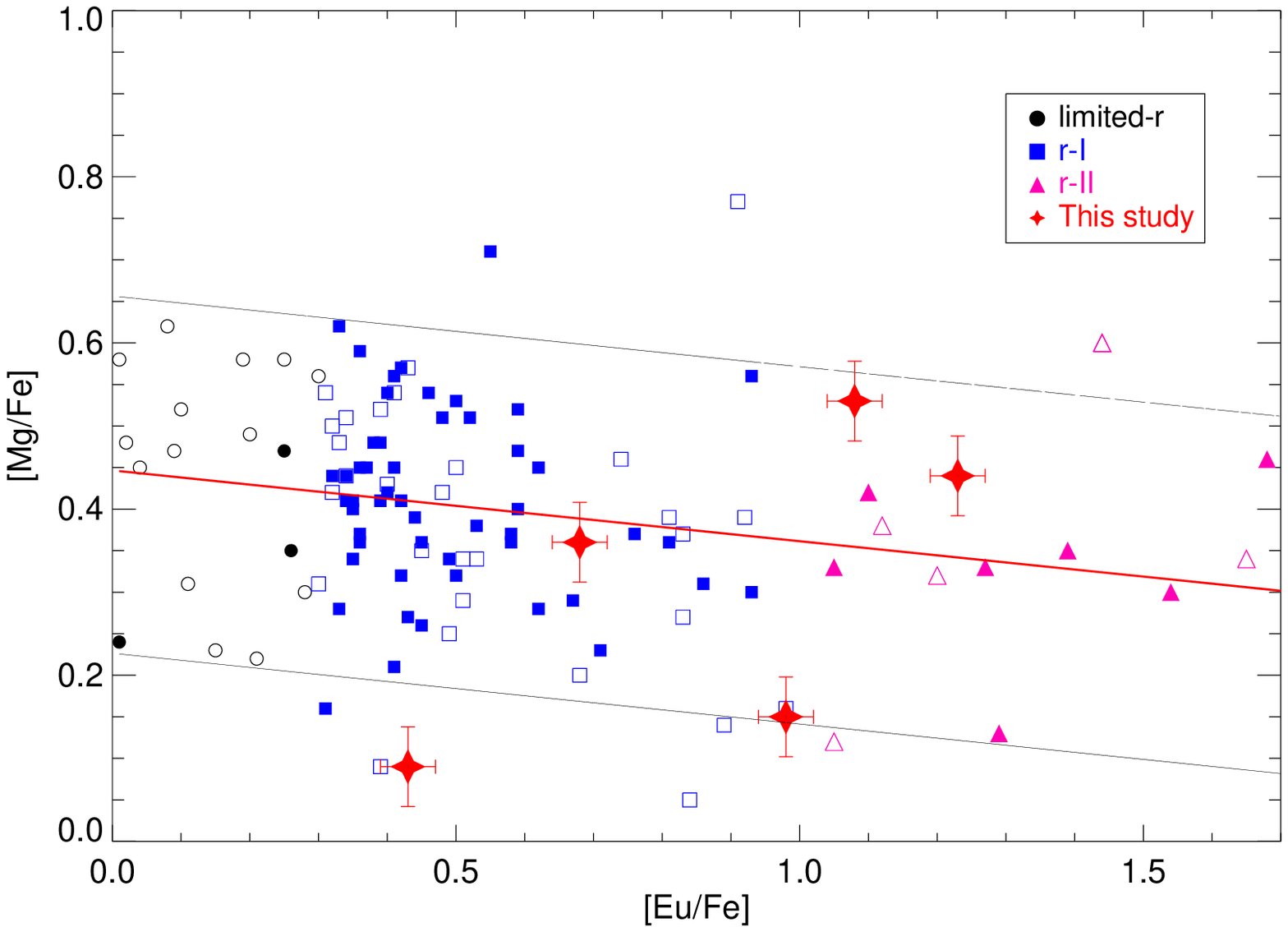}
\includegraphics[width=1.05\columnwidth]{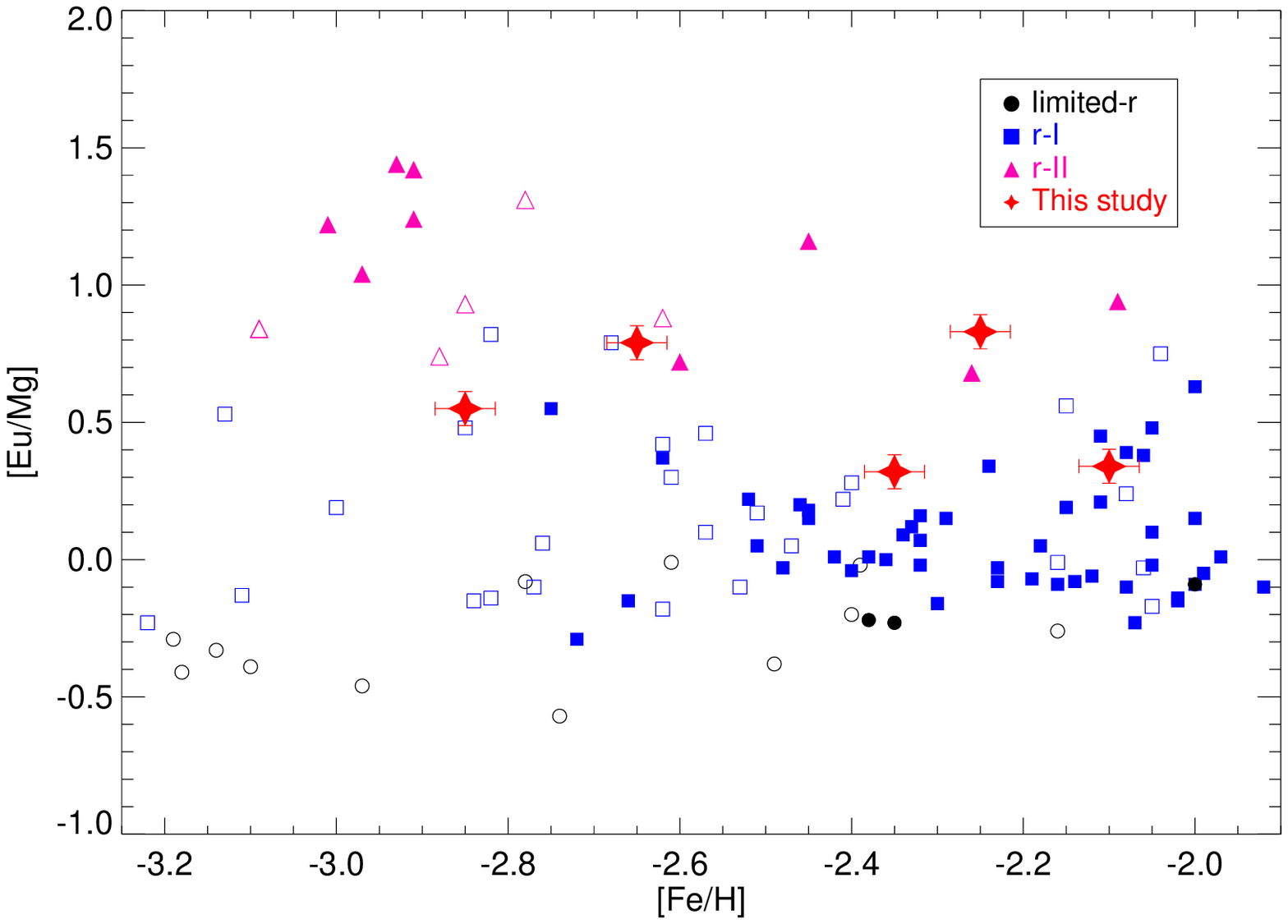}
\caption{ (Left panel) Distribution of [Mg/Fe], as a function of [Eu/Fe], for RPE stars. The program stars are
marked with filled red diamonds. The data for $r$-process-rich stars with measured Mg and Eu abundances is 
compiled  from various abundance studies of metal poor stars including \citet{mcwilliam1995,cayrel2004,
barklem2005,wako2005,hayek2009,roederer2014,jifrebelnature16,roederer16,sakari18rpa}. The open symbols 
represent data from the high resolution studies of metal poor stars without any biases in their target 
selection whereas the filled symbols represent the stars from studies aimed at $r$-process enhanced stars. 
The red solid line indicates the best fit of the data. 
The slope and $\sigma$ of the fit are $-0.09$ and 0.43 dex, respectively. The black lines indicate the 
1$\sigma$ width of the best fit. (Right panel) Distribution of [Eu/Mg], as a function of [Fe/H], for these 
same stars. }  
\end{figure*}

Figure 7 (left panel) shows the variation of Mg, which is primarily produced by Type-II SNe, as a function
of Eu, which is primarily produced by the $r$-process, for different classes of RPE stars. The data are 
compiled from the high-resolution surveys of metal-poor stars with an unbiased target selection which were 
not aimed at studying certain peculiar chemical signatures (e.g \citealt{mcwilliam1995,cayrel2004,wako2005,
roederer2014}) and are represented by open symbols in the figure. The filled symbols represent the data from 
the high resolution surveys for the RPE stars (e.g \citealt{hayek2009,jifrebelnature16,roederer2016ret,
sakari18rpa}). Both the samples (unbiased and 
RPE surveys) show the same trend with similar value of slope and $\sigma$. The $r$-II stars are found to be 
relatively Mg poor with a larger scatter, while limited-$r$ stars are Mg rich. The decrease in [Mg/Fe] is 
accompanied with an increase in [Eu/Fe]. The downward trend, as shown by the red solid line, is indicative 
of the decreasing contribution of SNe-II as $r$-process enrichment increases. More than 90\% of the sample 
falls within the 1$\sigma$ width from the average value of the fit marked by black lines. We conclude that 
limited-$r$ and $r$-I stars have received greater contributions from SNe-II compared to $r$-II stars. 
However, we note the existence of a significant scatter, and more data will be essential to carry out a 
detailed analysis on the significance of the slope. The right-hand panel of Figure 7 shows the [Eu/Mg] 
ratio as a function of [Fe/H].

\begin{figure*}[!htbp]
\centering
\includegraphics[width=1.25\columnwidth]{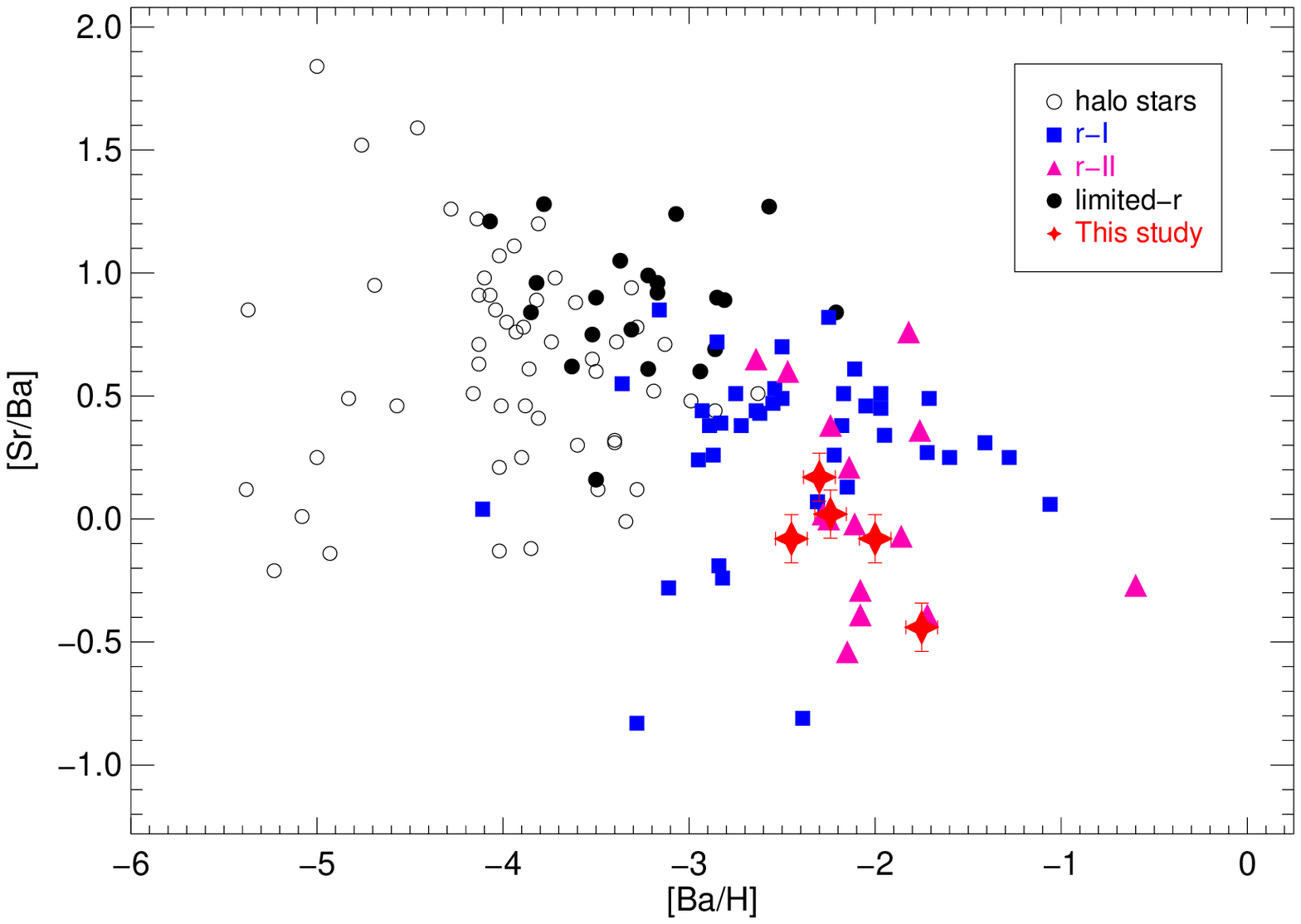}
\caption{Distribution of the RPE population in the [Ba/H] vs. [Sr/Ba] plane. The RPE population is found to 
contain the expected higher content of [Ba/H]. The data is compiled using the JINAbase \citep{jinabase} and 
SAGA databases \citep{sudasaga}, based on data from \citet{cayrel2004,honda2004,barklem2005, roederer2014, 
cain18, hansen18rpa}.
}
\end{figure*}

 Strontium is a surrogate element for the light $r$-process elements, and could be produced by both core-collapse supernovae (CCSNe), as well as by neutron star-neutron star mergers (NSMs), whereas the heavier element Ba is likely to only be produced by NSMs (or collapsars), as discussed by, e.g., \citet{tsuji1,tsuji2}, \citet{susmitha}, and \citet{siegel2019}.  Figure 8 shows the distribution of [Sr/Ba], as a function of [Ba/H], for the different classes of RPE and non-RPE, C-normal halo stars. The RPE stars, as expected, congregate towards the right side of the figure, with a higher abundance of [Ba/H], but the different classes of RPE stars also exhibit a mixed distribution. Both limited-$r$ and $r$-I stars exhibit a linear trend with respect to [Ba/H], however, the $r$-II stars show a more vertical distribution with little trend. This could indicate that the trend is due to dilution in the case of the limited-$r$ and $r$-I stars.  However, the $r$-II stars that do not exhibit higher abundances of heavier $r$-process elements (Ba in this case) is surprising, as they are expected to have undergone maximum $r$-process enrichment. For the $r$-II stars, the Ba distribution also spans a narrow region, and is more likely to have its origin from one enhancement episode, whereas the linear trend for the limited-$r$ and $r$-I stars indicate an evolutionary process.  The outlier among the $r$-II stars shown in Figure 8 is J18024226–4404426 \citep{placco2018aj} . It is also a metal rich r-II star with [Fe/H] = $-1.55$ and [Eu/Fe] = +1.05 \citep{hansen18rpa}. A relatively high [Ba/Eu] ratio of ($-$0.10) is found in this star, which might indicate a contribution from the $s$-process to the neutron-capture-element abundances at such high metallicity. However, the neutron-capture-element abundance pattern of this star is dominated by the r-process \citep{hansen18rpa} . J18024226–4404426 also widens the otherwise narrow metallicity distribution of the well studied r-II stars in the halo. Apart from the higher [Ba/H], the RPE stars are also found to have normal [Sr/Ba], which indicates that neither Sr or Ba are produced preferentially over on another. Thus, they likely have similar contributions from CCSNE or NSMs, and neither could be argued as the sole origin for metal-poor RPE stars from this evidence.

\subsection{A New Actinide-Boost Star}

 The abundance pattern of the light elements (such as C), the $\alpha$-elements (such as Mg), and the Fe-peak 
 elements (e.g., Fe, Co, Ni) in SDSS J0043+1948 are very similar to typical VMP halo stars. The abundance 
 pattern for the elements we could measure is shown in Figure 9, along with the other program stars.  Among 
 the light neutron-capture elements, Sr, Y, and Zr could be measured for this star; they exhibit a level and 
 spread that could be attributed to a limited $r$-process.


 The heavy neutron-capture elements for this star exhibit a lower degree of scatter, and compare well with 
 the scaled-Solar abundances (and other RPE stars from the literature). Unfortunately, due to the poor SNR 
 towards the blue end of the spectrum, among the elements approaching the second $r$-process peak, we could 
 only derive abundances for Ba, La, Ce, Nd, Sm, Eu, and Dy. They exhibit a consistent level of enhancement 
 (see Table 3), as expected in the main $r$-process.

\begin{figure*}[!htbp]
\centering
\includegraphics[width=1.5\columnwidth]{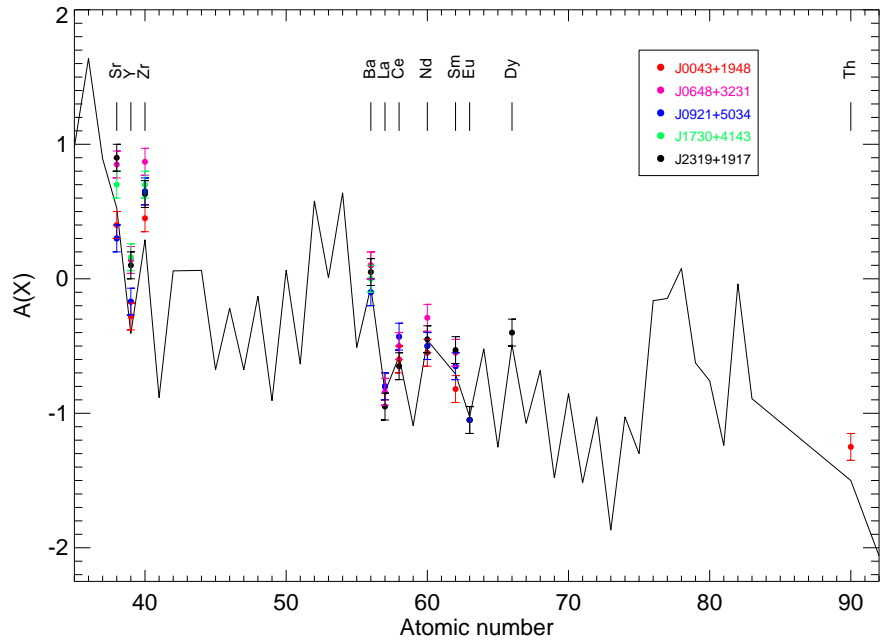}
\caption {The abundance pattern of the detected neutron-capture elements for our program stars. The gray 
line indicates the scaled-Solar $r$-process abundances; the derived abundances for the individual elements, 
normalized to the  Eu abundance, are indicated with different colors. The Solar $r$-process abundances were 
kindly provided by Dr. Erica Holmbeck (private communication), using the values computed by 
\citet{arlandini1999}.}
\end{figure*}

\begin{figure*}[!htbp]
\centering
\includegraphics[width=1.25\columnwidth]{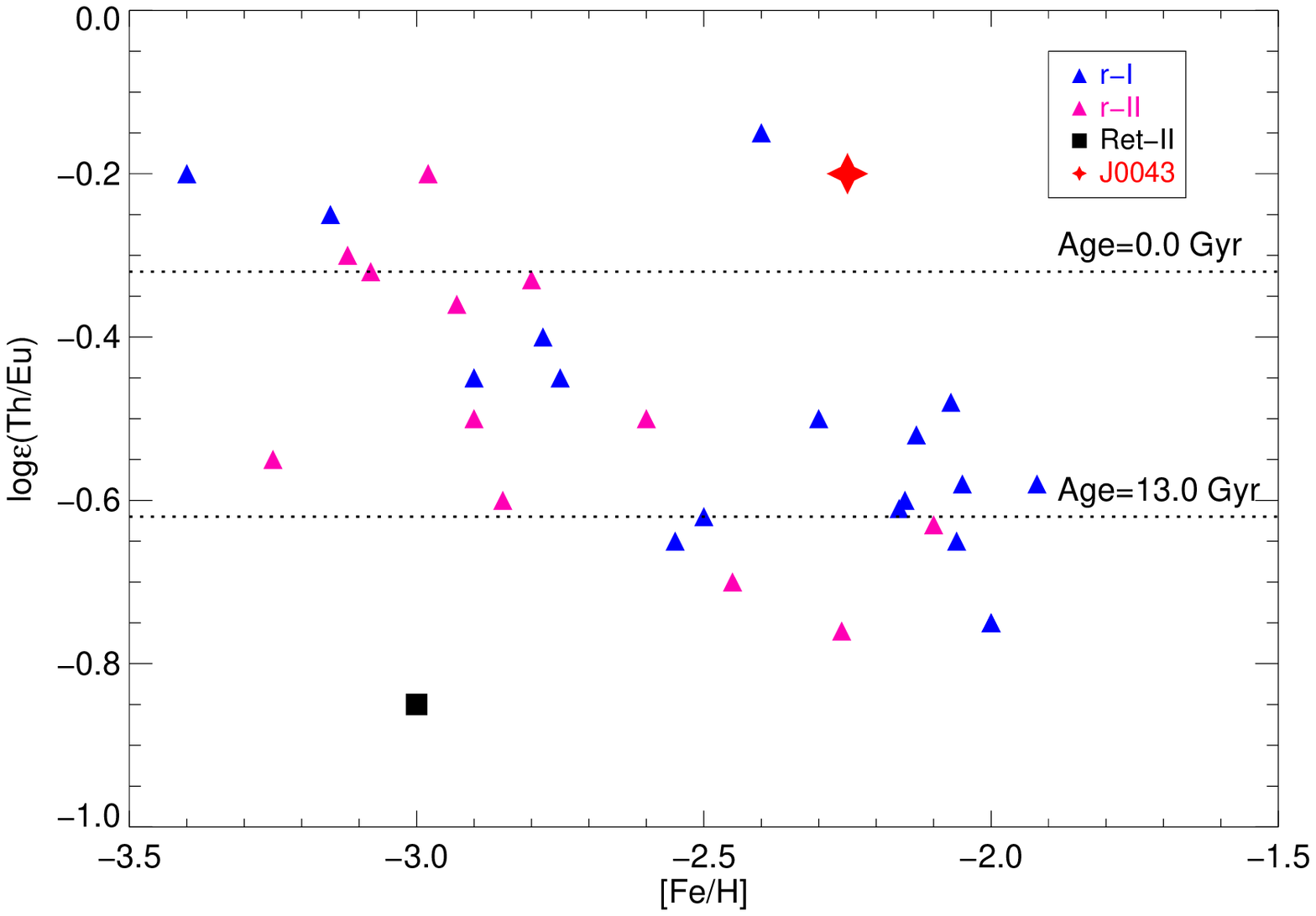}
\caption{ Distribution  of log $\epsilon$ (Th/Eu) for metal-poor stars with available measurements of Th, as 
a function of [Fe/H]. The program star SDSS J~0043+1948 is marked with a filled red diamond. Previously 
recognized $r$-I (blue triangles) and $r$-II (pink triangles) are shown for comparison, compiled from the 
RPA papers and other sources \citep{roederer2014,hill2017}. The black square shows the bright 
$r$-II star DES J033523−540407 in the Reticulum II galaxy \citep{jifrebel2018}. The dashed lines 
mark the corresponding ages, derived as in \cite{schatz2002} and  \citet{holmbeck18}. The stars with 
unusually large Th abundances, relative to Eu, the so-called actinide-boost stars, occupy the upper 
region of this diagram, and are associated with un-physical ages (as derived from their [Th/Eu] ratios).
} 
\end{figure*}

 Although thorium is one of the more difficult elements to measure in a stellar spectrum, we could measure 
 its abundance from the Th II line at 4019\,{\AA} for SDSS J0043+1948, obtaining a value [Th/Fe] = +1.28. 
 The synthesis of the Th II line  for this star is shown in the bottom-right panel of Figure 3. The measured 
 [Th/Eu] = +0.30 indicates a high actinide-to-lanthanide ratio, resulting in its classification as an 
 actinide-boost star.  The log $\epsilon$(Th/Eu) = $-$0.20  is higher than the ``normal" (non-actinide-boost) 
 stars, which typically exhibit values around $-$0.50 (see, e.g., \citealt{holmbeck18, holmbecknsm, holmbeck19} ).
A comparison of SDSS J0043+1948 with other RPE stars with available Th abundances is shown in Figure 10. 
The presence of the actinide boost for a significant number of stars demonstrates a large variation in the 
production of actinides. However, unlike the stars with scatter in the first $r$-process peak (Sr, Y and Zr), 
it is difficult to decouple the origin of actinide-boost stars from the main $r$-process production sites 
\citep{holmbeck18}. Measurement of the actinides Th and U for larger numbers of RPE stars are required to 
better probe the origin of this enhancement phenomenon. Mixing models as a diagnostic of the relative 
contributions of tidal debris and jet ejecta from a NSM, as discussed by \cite{holmbecknsm}, may be a 
promising avenue of exploration. 

 Detection and measurement of the actinides can also be used for  
the determination of age limits. using the technique of radioactive decay dating (e.g., \citealt{placco17}, 
and references therein). This appears to be true even for the actinide-boost stars.   Since U could not be 
measured for SDSS J0043+1948, only the Th/Eu chronometer could be used, employing the production ratios from 
\citep{schatz2002}. However, for this, and other actinide-boost stars, using the ratio of Th/Eu leads to 
un-physical age estimates, as the measured ratios are higher than the theoretically obtained initial 
production ratios (resulting in negative ages). The existing models for $r$-process nucleosynthesis fail 
to successfully derive the yields of actinide-boost stars. Measurement of U for this star (which will require
higher-resolution, and higher-SNR spectra of this star) would enable use of the Th/U chronometer, as both of 
these elements are likely to be affected by the actinide-boost proportionately. 

\section{Conclusions} 

 In this work, we have derived the detailed chemical abundances for five newly identified relatively bright 
 RPE stars, which are ideal candidates for detailed higher-resolution abundance studies, even with 
 modest-aperture telescopes. We have also compared our program stars with the different known classes 
 of RPE stars, and argue that the contribution from SNe-II decreases with the level of $r$-process 
 enrichment. We have also shown that the abundance for heavier $r$-process elements is enhanced for 
 the RPE stars, with $r$-II stars not following the usual trend as other halo RPE stars. The 
 $r$-process-element abundances of our program stars match with the scaled-Solar abundances, 
 conforming with the known universality of the main $r$-process, along with the usual scatter 
 for the lighter $r$-process elements. One of our program stars, SDSS~J0043+1948, is one of the 
 brightest known RPE stars with a Th measurement, and has been shown to be an actinide-boost star.
 This new sample of bright RPE stars offers the unique possibility for the detection of many more key 
 neutron-capture elements, from the ground and from space with HST.

\acknowledgments   We thank the anonymous referee for comments and suggestions that greatly helped to 
improve both the quality and presentation of our results.  We also thank the staff of IAO, Hanle and CREST, 
Hosakote, that made these observations possible. The facilities at IAO and CREST are operated by the Indian 
Institute of Astrophysics, Bangalore. We also thank Erica Holmbeck for providing the Solar $r$-process 
residuals in a private communication.  T.C.B. acknowledges partial support for this work from grant 
PHY 14-30152; Physics Frontier Center/JINA Center for the Evolution of the Elements (JINA-CEE), awarded 
by the US National Science Foundation.

\begin{table*}
\tabcolsep3.0pt
\begin{center}
\caption{Elemental-Abundance Determinations for \sdssfortythree} \label{c5t3}
\begin{tabular}{crrrrrrrrrrr}
\hline\hline
Element &Species & $N_{lines}$ & $A$(X) & Solar & [X/H] & [X/Fe] & $\sigma$ \\
\hline
C &CH &    &5.25 &8.43 &$-$3.18 &$-$0.93 &0.13 \\
Na &Na I &2 &4.05 &6.24 &$-$2.19 &$+$0.06 &0.17 \\
Mg &Mg I &5 &5.50 &7.60 &$-$2.10 &$+$0.15 &0.22 \\
Al &Al I &1 &2.96 &6.45 &$-$3.49 &$-$0.24 &0.34 \\
Ca &Ca I &5 &4.50 &6.34 &$-$1.84 &$+$0.41 &0.19\\
Sc &Sc II &4 &1.10 &3.15 &$-$2.15 &$+$0.10 &0.18\\
Ti &Ti I &4 &3.20 &4.95 &$-$1.75 &$+$0.50 &0.21\\
   &Ti II &3 &3.10 &4.95 &$-$1.85 &$+$0.40 &0.24\\
Cr &Cr I &3 &3.10 &5.64 &$-$2.54 &$-$0.29 &0.27\\
   &Cr II &2 &3.85 &5.64 &$-$1.81 &$+$0.44 &0.32\\
Mn &Mn I &3 &2.70 &5.43 &$-$2.73 &$-$0.48 &0.29\\
Co &Co I &2 &2.60 &4.99 &$-$-2.39 &$-$0.14 &0.24\\
Ni &Ni I &3 &3.90 &6.22 &$-$2.32 &$-$0.07 &0.27 \\
Cu &Cu I &1 &1.49 &4.19 &$-$2.70 &$-$0.45 &0.23 \\
Zn &Zn I &2 &2.60 &4.56 &$-$1.96 &$+$0.29 &0.22 \\
Sr &Sr II &2 &0.70 &2.87 &$-$2.17 &$+$0.08 &0.17\\
Y &Y II &2 &0.00 &2.21 &$-$2.21 &$+$0.04 &0.19 \\
Zr &Zr II &1 &0.75 &2.58 &$-$1.68 &$+$0.57 &0.21 \\
Ba &Ba II &2 &0.40 &2.18 &$-$1.78 &$+$0.47 &0.16 \\
La &La II &2 &$-$0.50 &1.11 &$-$1.61 &$+$0.64 &0.27 \\
Ce &Ce II &1 &$-$0.30 &1.58 &$-$1.88 &$+$0.37 &0.23 \\
Nd &Nd II &1 &$-$0.25 &1.42 &$-$1.67 &$+$0.58 &0.31 \\
Sm &Sm II &2 &$-$0.54 &0.96 &$-$1.60 &$+$0.75 &0.20 \\
Eu &Eu II &1 &$-$0.75 &0.52 &$-$1.27 &$+$0.98 &0.19 \\
Dy &Dy II &1 &$-$0.50 &1.10 &$-$1.60 &$< +$0.65 & 0.29 \\
Th &Th II &1 &$-$0.95 &0.02 &$-$0.97 &$+$1.28 &0.33 \\
\hline
\end{tabular}
\end{center} 
$\sigma$ indicates the total error for SNR, $T_{\rm eff}$, and $\log (g)$, added in quadrature.

\end{table*}

\begin{table*}
\tabcolsep3.0pt
\begin{center}
\caption{Elemental-Abundance Determinations for \sdsssixfortyeight} \label{c5t4}
\begin{tabular}{crrrrrrrrrrr}
\hline\hline
Element &Species & $N_{lines}$ & $A$(X) & Solar & [X/H] & [X/Fe] & $\sigma$ \\
\hline
C &CH &    &5.50 &8.43 &$-$2.93 &$-$0.58 &0.15 \\
Na &Na I &2 &4.11 &6.21 &$-$2.10 &$+$0.25 &0.21 \\
Mg &Mg I &5 &5.60 &7.59 &$-$1.99 &$+$0.36 &0.28 \\
Al &Al I &1 &2.41 &6.43 &$-$4.02 &$-$0.67 &0.31\\
Ca &Ca I &8 &4.43 &6.32 &$-$1.89 &$+$0.46 &0.18\\
Sc &Sc II &3 &0.89 &3.15 &$-$2.26 &$+$0.09&0.17\\
Ti &Ti I &7 &3.07 &4.93 &$-$1.86 &$+$0.49 &0.19\\
   &Ti II &9 &2.90 &4.93 &$-$2.03 &$+$0.32 &0.22\\
Cr &Cr I &5 &3.32 &5.62 &$-$2.30 &$+$0.05 &0.24\\
   &Cr II &3 &3.54 &5.62 &$-$2.08 &$+$0.27 &0.27\\
Mn &Mn I &2 &2.77 &5.42 &$-$2.65 &$-$0.30 &0.26\\
Co &Co I &2 &2.68 &4.93 &$-$2.25 &$+$0.10 &0.23\\
Ni &Ni I &4 &4.32 &6.20 &$-$1.88 &$+$0.47 &0.21 \\
Cu &Cu I &1 &1.03 &4.19 &$-$3.16 &$+$0.81 &0.30 \\
Zn &Zn I &2 &2.44 &4.56 &$-$2.12 &$+$0.23 &0.23 \\
Sr &Sr II &2 &0.75 &2.83 &$-$2.08 &$+$0.27 &0.18\\
Y &Y II &2 &0.00 &2.21 &$-$2.21 &$+$0.14 &0.21 \\
Zr &Zr II &1 &0.75 &2.59 &$-$1.84 &$+$0.51 &0.20 \\
Ba &Ba II &2 &0.25 &2.25 &$-$2.00 &$+$0.35 &0.17 \\
La &La II &3 &$-$0.94 &1.11 &$-$2.05 &$+$0.30 &0.22 \\
Ce &Ce II &2 &$-$0.60 &1.58 &$-$2.18 &$+$0.17 &0.24 \\
Nd &Nd II &3 &$-$0.35 &1.42 &$-$1.77 &$+$0.33 &0.26 \\
Sm &Sm II &2 &$-$0.65 &0.96 &$-$1.61 &$+$0.74 &0.22 \\
Eu &Eu II &1 &$-$1.15 &0.52 &$-$1.67 &$+$0.68 &0.21 \\
\hline
\end{tabular}
\end{center}
    $\sigma$ indicates the total error for SNR, $T_{\rm eff}$, and $\log (g)$, added in quadrature.
\end{table*}

\begin{table*}
\tabcolsep3.0pt
\begin{center}
\caption{Elemental-Abundance Determinations for \sdssninetwentyone} \label{c5t6}
\begin{tabular}{crrrrrrrrrrr}
\hline\hline
Element &Species & $N_{lines}$ & $A$(X) & Solar & [X/H] & [X/Fe] & $\sigma$ \\
\hline
C &CH &    &5.50 &8.43 &$-$2.93 &$-$0.28 &0.16 \\
Na &Na I &2 &4.08 &6.21 &$-$2.13 &$+$0.52 &0.20 \\
Mg &Mg I &4 &5.38 &7.59 &$-$2.21 &$+$0.44 &0.25 \\
Al &Al I &1 &2.39 &6.443 &$-$4.04 &$-$0.39 &0.33\\
Ca &Ca I &7 &4.13 &6.32 &$-$2.19 &$+$0.46 &0.21\\
Sc &Sc II &1 &0.28 &3.15 &$-$2.87 &$-$0.22 &0.19\\
Ti &Ti I &6 &2.42 &4.93 &$-$2.51 &$+$0.14 &0.25\\
   &Ti II &6 &2.65 &4.93 &$-$2.28 &$+$0.37 &0.28\\
Cr &Cr I &5 &2.95 &5.62 &$-$2.67 &$-$0.02 &0.24\\
   &Cr II &1 &3.57 &5.62 &$-$2.05 &$+$0.60 &0.32\\
Mn &Mn I &4 &2.04 &5.42 &$-$3.38 &$-$0.73 &0.26\\
Co &Co I &2 &2.06 &4.99 &$-$2.87 &$-$0.22 &0.24\\
Ni &Ni I &3 &4.18 &6.20 &$-$2.02 &$+$0.63 &0.27 \\
Zn &Zn I &2 &2.44 &4.56 &$-$2.12 &$+$0.53 &0.22 \\
Sr &Sr II &2 &0.55 &2.83 &$-$2.28 &$+$0.37 &0.21\\
Y &Y II &2 &0.00 &2.21 &$-$2.21 &$+$0.44 &0.22 \\
Zr &Zr II &1 &0.80 &2.59 &$-$1.79 &$+$0.86 &0.18 \\
Ba &Ba II &2 &0.05 &2.25 &$-$2.25 &$+$0.40 &0.17 \\
La &La II &3 &$-$0.65 &1.11 &$-$1.76 &$+$0.89 &0.21\\
Ce &Ce II &2 &$-$0.25 &1.58 &$-$1.83 &$+$0.82 &0.24 \\
Nd &Nd II &4 &$-$0.40 &1.42 &$-$1.82 &$+$0.83 &0.22 \\
Sm &Sm II &2 &$-$0.50 &0.96 &$-$1.46 &$+$1.19 &0.26 \\
Eu &Eu II &1 &$-$0.90 &0.52 &$-$1.42 &$+$1.23 &0.23 \\
Dy &Dy II &1 &$-$0.15 &1.10 &$-$1.25 &$<$1.40 &0.34 \\
\hline
\end{tabular}
\end{center}

       $\sigma$ indicates the total error for SNR, $T_{\rm eff}$, and $\log (g)$, added in quadrature.
    
\end{table*}

\begin{table*}
\tabcolsep3.0pt
\begin{center}
\caption{Elemental Abundance Determinations for \sdssseventeenthirty} \label{c5t7}
\begin{tabular}{crrrrrrrrrrr}
\hline\hline
Element &Species & $N_{lines}$ & $A$(X) & Solar & [X/H] & [X/Fe] & $\sigma$ \\
\hline
C &CH &    &5.50 &8.43 &$-$2.93 &$-$0.08 &0.21 \\
N &CN &    &6.00 &7.83 &$-$1.83 &$+$1.02 &0.34\\
Na &Na I &2 &3.40 &6.21 &$-$2.81 &$+$0.04 &0.19 \\
Mg &Mg I &3 &5.30 &7.59 &$-$2.29 &$+$0.56 &0.24 \\
Al &Al I &1 &2.55 &6.43 &$-$3.88 &$-$0.03 &0.33\\
Ca &Ca I &9 &3.84 &6.32 &$-$2.48 &$+$0.37 &0.20\\
Sc &Sc II &2 &$-$0.30 &3.15 &$-$3.45 &$-$0.60 &0.22\\
Ti &Ti I &8 &2.60 &4.93 &$-$2.33 &$+$0.52 &0.24\\
   &Ti II &4 &2.20 &4.93 &$-$2.73 &$+$0.12 &0.26\\
Cr &Cr I &4 &2.80 &5.62 &$-$2.82 &$+$0.03 &0.21\\
   &Cr II &1 &3.00 &5.62 &$-$2.62 &$+$0.23 &0.28\\
Ni &Ni I &4 &3.60 &6.22 &$-$2.62 &$+$0.23 &0.27 \\
Cu &Cu I &1 &0.87 &4.19 &$-$3.32 &$+$0.47 &0.30 \\
Zn &Zn I &1 &2.41 &4.56 &$-$2.15 &$+$0.70 &0.23 \\
Sr &Sr II &2 &0.50 &2.83 &$-$2.33 &$+$0.52 &0.21\\
Y &Y II &2 &0.00 &2.21 &$-$2.21 &$+$0.64 &0.24 \\
Zr &Zr II &1 &0.50 &2.58 &$-$2.08 &$+$0.77 &0.24 \\
Ba &Ba II &2 &$-$0.20 &2.25 &$-$2.45 &$+$0.40 &0.21 \\
Eu &Eu II &1 &$-$1.25 &0.52 &$-$1.77 &$+$1.08 &0.25 \\
\hline
\end{tabular}
\end{center}

       $\sigma$ indicates the total error for SNR, $T_{\rm eff}$, and $\log (g)$, added in quadrature.
       
  \end{table*}

\begin{table*}
\tabcolsep3.0pt
\begin{center}
\caption{Elemental Abundance Determinations for \sdsstwentythree} \label{c5t9}
\begin{tabular}{crrrrrrrrrrr}
\hline\hline
Element &Species & $N_{lines}$ & $A$(X) & Solar & [X/H] & [X/Fe] & $\sigma$ \\
\hline
C &CH &    &5.75 &8.43 &$-$2.68 &$-$0.58 &0.17 \\
Na &Na I &2 &4.12 &6.21 &$-$2.09 &$+$0.01 &0.21\\
Mg &Mg I &5 &5.58 &7.59 &$-$2.01 &$+$0.09 &0.24 \\
Al &Al I &1 &2.42 &6.43 &$-$4.01 &$-$0.91 &0.38\\
Si &Si I &1 &5.01 &7.51 &$-$2.50 &$-$0.40 &0.28 \\
Ca &Ca I &8 &4.17 &6.32 &$-$2.15 &$-$0.05 &0.19\\
Sc &Sc II &3 &0.82 &3.15 &$-$2.33 &$-$0.23 &0.22\\
Ti &Ti I &7 &2.89 &4.93 &$-$2.04 &$+$0.06 &0.21\\
   &Ti II &6 &2.63 &4.93 &$-$2.30 &$-$0.20 &0.26\\
Cr &Cr I &5 &3.35 &5.62 &$-$2.27 &$-$0.17 &0.25\\
   &Cr II &2 &3.1 &5.62 &$-$2.52 &$-$0.42 &0.29\\
Mn &Mn I &1 &2.70 &5.42 &$-$2.72 &$-$0.62 &0.27\\
Co &Co I &2 &2.26 &4.93 &$-$2.67 &$-$0.57 &0.28\\
Ni &Ni I &4 &4.45 &6.20 &$-$1.75 &$+$0.35 &0.26 \\
Zn &Zn I &1 &2.27 &4.56 &$-$2.29 &$-$0.19 &0.23 \\
Sr &Sr II &2 &0.80 &2.83 &$-$2.03 &$+$0.07 &0.25\\
Y &Y II &2 &0.00 &2.21 &$-$2.21 &$-$0.11 &0.23 \\
Zr &Zr II &1 &0.50 &2.59 &$-$2.09 &$+$0.01 &0.27 \\
Ba &Ba II &2 &$-$0.05 &2.25 &$-$2.30 &$-$0.20 &0.22 \\
La &La II &3 &$-$1.15 &1.11 &$-$2.26 &$-$0.16 &0.23 \\
Ce &Ce II &3 &$-$0.80 &1.58 &$-$2.38 &$-$0.28 &0.20 \\
Nd &Nd II &5 &$-$0.50 &1.42 &$-$1.92 &$+$0.18 &0.25 \\
Sm &Sm II &4 &$-$0.50 &0.96 &$-$1.46 &$+$0.64 &0.21 \\
Eu &Eu II &1 &$-$1.15 &0.52 &$-$1.67 &$+$0.43 &0.20 \\
Dy &Dy II &1 &$-$0.40 &1.10 &$-$1.50 &$+$0.60 &0.24 \\

\hline
\end{tabular}
\end{center}
      $\sigma$ indicates the total error for SNR, $T_{\rm eff}$, and $\log (g)$, added in quadrature.
\end{table*}

\begin{table*}
\begin{center}
\caption{Space velocities of the Program Stars}
\begin{tabular}{ccccccccccr}
\hline\hline
Object &U &V &W \\
       &(km/s) &(km/s) &(km/s) \\
\hline

SDSS~J0043+1948  &$-$126.83 &$-$173.36 &46.53\\
SDSS~J0648+2321 &108.08  &$-$335.83  &04.47 \\
SDSS~J0921+5034  &51.78  &$-$305.07 &$-$141.06\\
SDSS~J1730+4143 &215.23  &$-$81.99 &12.89 \\
SDSS~J2319+1917  &106.229 &$-$280.78 &97.60  \\

\hline
\end{tabular}
\end{center}

\end{table*}

\clearpage

\newpage
\end{document}